\documentclass[artical]{elsarticle}[2p]
\usepackage{lineno,graphicx,caption2}
\usepackage{float}
\usepackage[utf8]{inputenc}
\usepackage{multirow}
\usepackage[table,xcdraw]{xcolor}
\begin{document}
\begin{frontmatter}

\title{Community detection based on structural balance in signed networks}


\author[mymainaddress]{Peng Zhang}
\author[mymainaddress]{Xianyu Xu}
\author[mymainaddress,mysecondaryaddress]{Leyang Xue\corref{mycorrespondingauthor}}
\cortext[mycorrespondingauthor]{Corresponding author}
\ead{hsuehleyang@gmail.com}

\address[mymainaddress]{School of Science, Beijing University of Posts and Telecommunications, Beijing 100876, China}
\address[mysecondaryaddress]{International Academic Center of Complex Systems, Beijing Normal University, Zhuhai, 519087, China}

\begin{abstract}
In signed networks, some existing community detection methods treat negative connections as intercommunity links and positive ones as intracommunity links. 
However, it is important to recognize that negative links on real-world networks also play a key role in maintaining community stability. 
In this work, our aim is to identify communities that are not only densely connected but also harmonious or balanced in terms of the nature of their relationships. 
Such communities are more likely to be stable over time and less prone to conflicts.
Consequently, we propose a motif-based method to identify communities by quantifying the importance of links in the local structural balance.
The results in synthetic and real-world networks show that the proposed method has a higher performance in identifying the community. 
In addition, it demonstrates strong robustness, i.e., remains insensitive to the balance of the network, and accurately classifies communities in real-world networks.
\end{abstract}

\end{frontmatter}

\section*{Introduction}
Social networks allow users to participate in online activities, freely express their emotions, and generate a wealth of emotional information in user interactions, including liking and giving thumbs up to followers~\cite{jin2013understanding, maia2008identifying}.
Relationships among users in online networks often display not only positive emotions, but also intense negative ones~\cite{leskovec2010signed}.
As user emotions are intricately linked to their social connections, this information can provide new insight into many practical issues such as information dissemination~\cite{zhang2022impact}, evolution of opinions~\cite{li2022modeling}, and social cooperation~\cite{zhong2021evolution}.
To quantitatively describe such information, signed networks provide a math tool where positive links represent "consensus" or "friendship", while negative links represent "opposition" or "hostility".
Due to its generality, signed networks have been widely used to describe various real systems to solve specific tasks~\cite{gu2022signed,zhang2019new, daud2020applications}.
For example, positive and close relationships between users on on-line systems, indicating shared interests and hobbies, offer new perspectives for item recommendations~\cite{zhang2019new}. 
In addition to the aforementioned applications, the utilization of sign information to detect community structures has recently received considerable attention~\cite{yang2007community,anchuri2012communities, esmailian2015community,sun2020stable}.

Currently, researchers propose community detection algorithms for signed networks from a variety of perspectives that can roughly be divided into two categories.
The first uses iterative heuristic algorithms to minimize frustration functions.
Patrick et al. presented a method (DM, the method proposed by Doreian and Mrvar) to detect the community~\cite{doreian1996partitioning}, in which a new function has been developed to minimize the number of negative links within the community and the positive links between communities.
Yang et al. proposed a finding and extraction community (FEC) algorithm, which incorporates sign and density information into the clustering coefficients and finally finds communities by optimizing the target functions~\cite{doreian1996partitioning}.
Tushar et al. used the cluster re-clustering algorithm (CRA) to detect community, which was expanded based on a breadth-first clustering algorithm~\cite{sharma2009community}.
The second is to introduce the modularity concept into signed networks, then redefine the modularity function by taking into account positive and negative information.
Sepecifically, VA et al. introduced modularity in the Potts model and redefined it to detect community structures in signed networks~\cite{traag2009community}.
Li et al. combined modularity with the GN method to design GN-H co-training algorithms to find the community using emotional and social information~\cite{li2010exploiting}. 
Guo et al. added link sign information to modularity and combined it with random walking to detect small communities~\cite{guo2015research}. 

The above work detects communities based on the objective that each node is assigned to a community such that the most positive links appear within the communities and the majority of negative links occur between communities. 
However, such an objective might be idealised, especially on social networks where negative relationships within communities are common~\cite{festinger1950social}.  
Here, we generalise the concept of community in signed networks, focusing on community formation.
Research suggests that a network's balance indicates its evolutionary stage: as it matures, its balance strengthens~\cite{pan2018structural}. 
As individuals in social networks establish relationships, these ties evolve over time. 
A naturally formed community should have a robust structural balance; otherwise, internal relationship imbalances can lead to its collapse or the disintegration of the community.
We believe that a community is not only a cluster of interconnected nodes, but also reflects the relative strength of the structure's balance among these nodes.

In this work, communities in signed networks are detected by the following two criteria:
(1) structural connective: The network is divided into groups where nodes are closely connected within each group but less connected between different groups.
(2) structural balance: The network is divided into subsets of nodes so that the structure within the group is more balanced.
Consequently, we propose a motif-based method to detect the community.
We use balanced motifs to quantify the role of links in local structural balance, transform the adjacency matrix into a weight matrix, and identify communities using a modularity function.

Section 2 describes the motif-based method and provides illustrative examples of community detection in signed networks.
In Section 3, we simulate numerical experiments on real and synthetic networks to test the performance of a motif-based method. 
We also present simulations on both real and synthetic networks, revealing the advantage of method over the DM algorithm, especially in synthetic networks with obvious community, irrespective of network balance. 
In real networks, the method accurately detects the benchmark communities.
In Section 4, we discuss the motif-based method in detail.

\section*{Method}
\textbf{The description of community detection on signed network}.
Take into account a signed network $G$ with $n$ nodes (e.g., $1,2,...,n$) and $m$ links, it is represented by the adjacency matrix $A$.
The elements of $A$ are assigned values based on the presence of a positive or negative link between node $i$ and node $j$. 
If there is a positive link, $A_{ij}$ is given a value of 1; if there is a negative link, $A_{ij}$ is given a value of -1; and if there is no link, $A_{ij}$ is given a value of 0.
We separate the adjacency matrix of the signed network into two components, $A^-$ and $A^+$. $A_{ij}^+$ is set to 1 if $A_{ij}$ is 1, and 0 otherwise. 
Similarly, $A_{ij}^-$ is set to 1 if $A_{ij}$ is -1 and 0 otherwise. 
Therefore, $A = A^+ - A^-$.
The links between node $i$ and node $j$ that are not directional, denoted by $e_{ij}$.
Our aim is to assign each node $i$ to one of the $c$ communities $\sigma_i\in \{1,2,...,c\}$ in a way that each community has the following characteristics:
(1) The balance of the community structure that is made up of various pairs of nodes is very strong.
(2) Each node has more links within the community than between the community.
Such a definition of communities is not as stringent as that one seeks to minimise the number of negative links within them.
In this work, detected communities are associated with stronger structural balance and connectivity, even when there are some negative links within the community.

\textbf{Motif-based community detection method}.
We propose an approach that uses motifs to discover the community in the signed network.
This approach assesses the importance of links based on their contribution to structural balance and assigns higher weights to links that play a significant role in network balance. Then, these links with higher weights are more likely to be assigned within communities.
In contrast, links that are not essential for maintaining the structural stability of networks are categorised as inter-community links.
Correspondingly, the nodes connected by these links are grouped into the same community.
Indeed, accurately measuring the significance of a link in the structural balance of networks can be challenging due to the higher complexity involved.
To simplify the problem, the concept is redefined to prioritise local structural balance over global structural balance. 
By analysing the importance of the link on the local structural balance, the evaluation becomes more practical and feasible.
Therefore, we quantify here the role of a link in the balance of all the motifs in which it is involved. 
If a link has a stronger influence on the local structural balance compared to what would be expected in a random network configuration, it is considered significant for the formation of the community. 
The modularity function is used to identify the community within the network.
The implementation of this approach mainly involves the following two parts.

\begin{figure}[!htbp]
	\centering
	\includegraphics[width =1.0 \textwidth]{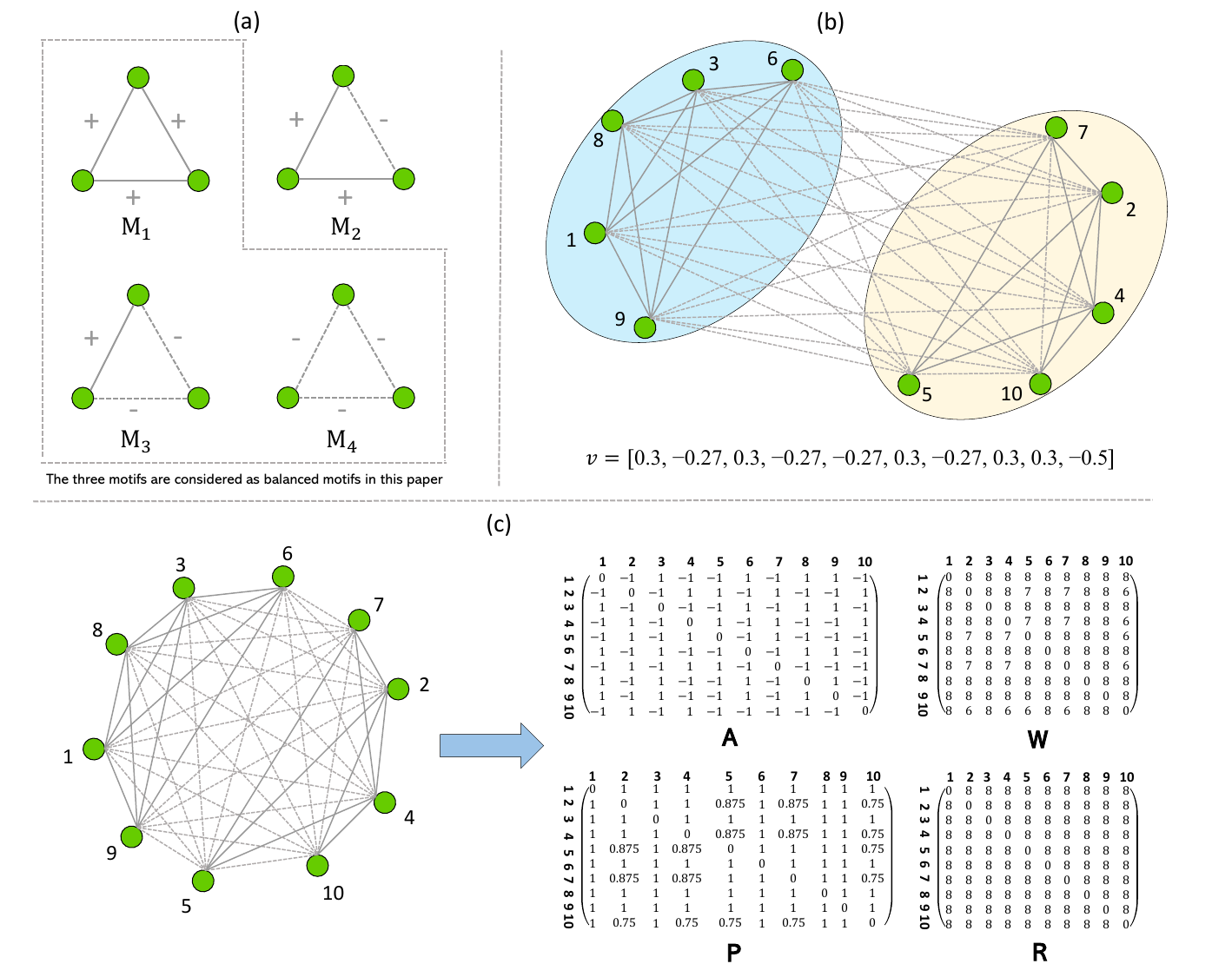}
	\caption{The illustration of motif-based approach to detect communities in a signed network.
		(a) Four basic motifs in signed network. 
		The balance motif is $M_1$, $M_3$, where $M_4$ is considered a weak balanced motif in social network. 
		(b) The result of community detection using the motif-based approach. 
		The motif-based approach measures the importance of each link in maintaining local structural balance by using balanced motifs (e.g. $M_1$, $M_3$, $M_4$), which assigns a numerical value to each link and maps the adjacency matrix ($A$) to the weight matrix ($P$). 
		The principle of identifying the community is that the weight of a link, reflecting its role in local structural balance, is greater than expected in a random configuration.
		Community detection is implemented by optimizing the weighted Q function, namely the eigenvector corresponding to the largest eigenvalue is determined using the modularity matrix.
		Finally, we obtain the positive and negative information of the corresponding positions of all nodes in the eigenvector and divide the nodes into distinct communities based on the same sign information.
		(c) We provide the matrix representation of $A$, $R$, $W$, and $P$ in the example where the solid and dashed lines, respectively, denote the positive and negative links.}
        \label{fig:1}
\end{figure}

\textbf{Quantifying the importance of a link in local structural balance}.
Triangles are fundamental building blocks of network connectivity, and their balance directly impacts the stability and dynamics of the network.
Furthermore, quantifying the importance of a link in the local balance of the structure formed by triangles is relatively straightforward. 
Since triangles have a simple and well-defined structure, it is easier to assess the impact of a link on the balance within these local configurations. 
According to early research on sociological networks, researchers have proposed four fundamental triangular structures in signed networks~\cite{heider1946attitudes, harary1953notion,cartwright1956structural, davis1967clustering}, as shown in Fig.~\ref{fig:1}(a).
Four triangles can be considered basic motifs, with $M_1$ and $M_3$ being structural balances since the product of the three links is positive.
However, such a definition of balance seems too stringent in social and biological networks.
Most real-world networks do not achieve complete structural balance and the balance observed in these networks is usually weak.
Davis proposed that the $M_4$ motifs demonstrate a weak balance~\cite{davis1967clustering}. 
Therefore, we take $M_4$ into account as one of the fundamental balanced motifs to enhance the practicality of actual networks.

A new metric $p$ is proposed to measure the importance of a link in the local structural balance,
\begin{equation}\label{1}
	p_{ij}={\frac{w_{ij}}{r_{ij}}},
\end{equation}
where $w_{ij}$ denotes the total number of times that the link $e_{ij}$ between node $i$ and node $j$ appears in three balanced motifs $M_1$, $M_3$ and $M_4$. 
Similarly, the total number of times the link $e_{ij}$ appears in all four motifs is denoted by $r_{ij}$.
When $e_{ij}$ forms a triangle with the other two links, that is, $r_{ij}\neq0$, then $p_{ij}$ make sense and $p_{ij} \in [0,1]$, otherwise we assign $p_{ij}$ = 0. 
The importance of a link in maintaining balance within the local structure increases as the value of $p_{ij}$ increases, and vice versa. 
To simplify the notation, the metric $p$ is expressed in a matrix form, represented by Eq.~\ref{eqs:eq2}.
\begin{equation}\label{eqs:eq2}
	{P=W./R},
\end{equation}
where the matrix $W$ and $R$ represent the number of times a link appears in the balanced motifs and in all motifs, respectively. 
The elements in both $W$ and $R$ are denoted by $w_{ij}$ and $r_{ij}$, respectively.
We determine the weight matrix $P$ in Eq.~\ref{eqs:eq2} by dividing the elements in the same position of the two matrices.
The values of $W$ and $R$ can be further calculated using Eqs.~\ref{eqs:eq3} and \ref{eqs:eq4}, respectively.
\begin{equation}\label{eqs:eq3}
		W=A^+ \cdot (A^+\times A^+)+A^-\cdot (A^-\times A^-)+A^-\cdot (A^-\times A^+)+\\A^-\cdot (A^+\times A^-)+A^+\cdot (A^-\times A^-),
\end{equation}

\begin{equation}\label{eqs:eq4}
	{R=(A^++A^-)\cdot(A^++A^-)\times(A^++A^-)},
\end{equation}
where $A^+$ and $A^-$ are, respectively, the positive and negative adjacency matrices. 
The symbol $\times$ is used for matrix multiplication, while $\cdot$ is used for numerical multiplication in the same position of two matrices.
By calculating the ratio of times that a link appears in balanced motifs to all motifs, we quantify the importance of links in local structural balance. 

We show an illustration example in Fig.~\ref{fig:1}(c).
The matrix $W$ and $R$ represent the frequency of each link appearing in the three balanced motifs (e.g. $M_1$, $M_3$ and $M_4$) and in all motifs, respectively.
We calculate the proportion of $W$ to $R$ and further obtain the matrix $P$, where $p_{ij}$ quantifies the importance of a link in local structural balance. 
For example, the link $e_{13}$ appearing in balanced motifs is as follows: $\triangle_{1 2 3 }$,$\triangle_{ 1 3 4}$,$\triangle_{1 3 5}$,$\triangle_{1 3 6}$,$\triangle_{1 3 7}$,$\triangle_{1 3 8}$,$\triangle_{1 3 9}$,$\triangle_{1 3 10}$, we have $w_{13} = 8$. 
The $e_{13}$ is present in 8 motifs and is not found in any imbalanced motif, so $r_{13} = 8$  and the final result $p_{13}=1$.

\textbf{Detecting the community structure}.
We employ a motif-based method to assess the importance of a link in the local structural balance, and further detect the community in signed networks. 
If the weight of a link between two nodes is higher than what would be expected in a random configuration, then it is more likely that the link is part of the community. 
We use the modularity function to identify communities in signed networks.
The function $Q$ is a global function that is used to measure the density difference between the real network and a null model~\cite{blondel2008fast,chang2011general,newman2006modularity}. 
It quantifies the extent to which the network's community structure deviates from what would be expected in a random network.
A higher $Q$ value indicates a more pronounced community structure within the network.
Here, we use a weighted function $Q$ to detect communities within the network. 
Instead of using the traditional adjacency matrix, we replace it in the $Q$ function with the weight matrix $P$. 
Several related studies have demonstrated the feasibility of incorporating a weight matrix into the $Q$ function~\cite{newman2004analysis,fan2007accuracy}. 
When incorporating the weights of the links, the modified function $Q$ can effectively capture the connective strength and balance influence of these links, leading to improved results in identifying meaningful communities within the network.
The weighted function $Q$ is defined as follows:
\begin{equation}\label{eqs:eq5}
	{Q}=\frac{1}{w}\sum_{ij}(p_{ij}-\frac{p_ip_j}{w})\delta(c_i,c_j),
\end{equation}
\begin{equation}\label{eqs:eq6}
	{h_{ij}=p_{ij}-\frac{p_ip_j}{w}},
\end{equation}
where the $h_{ij}$ is element of $H$, and $w$ denotes the overall weight of $P$.
The sum of the elements in the row $j$ of $P$ is represented by $p_j$, characterised as the extent to which node $j$ contributes to the balance structure.
Nodes $i$ and $j$ are represented by their respective communities $c_i$ and $c_j$.
When nodes $i$ and $j$ belong to the same community, $c_i = c_j$ and $\delta(c_i, c_j) = 1$. 
Conversely, if $c_i \neq c_j$, then $\delta(c_i, c_j) = 0$.
The modularity function is optimised to determine the eigenvector associated with the largest eigenvalue of the matrix $H$~\cite{newman2006finding}.
Finally, we extract the positive and negative information from the corresponding positions of all nodes in the eigenvector. 
Based on the sign of these values, we divide the nodes into distinct communities, grouping the nodes with the same sign information.

\section*{Experiments}
\subsection*{Datasets}
\textbf{Synthetic networks}.
We examine the performance of the motif-based method on different signed network configurations by creating a number of synthetic networks~(SN) as a benchmark to validate its effectiveness.
We construct a model to captures the structural interactions between communities by considering the probabilities of positive and negative links both between and within communities, using the parameters $p^{in}$, $p^+_{out}$, and $p^-_{in}$, as outlined in the following~\cite{huang2018snmfp}:
\begin{equation}\label{eqs:eq6}
	{SN(c,(n_1,n_2,...,n_c),k,p^{in},p^+_{out},p^-_{in})}
\end{equation}
The number of communities, $c$, is denoted by $n_1,n_2,...,n_c$, each of which has a degree of $k$. 
The $p^{in}$ controls the density of internal links within communities, $p^+_{out}$ determines the probability of positive links between communities, and $p^-_{in}$ governs the probability of negative links within communities.
In this simplified model, it is denoted $SN(c, n, k, p^{in}, p^+_{out}, p^-_{in})$, where the number of nodes is assumed to be equal within each community ($n_1=n_2=...=n_c$).

\textbf{Real-world networks}.
In addition to testing our method on synthetic networks, we also applied it to two real networks to identify the communities and assess its effectiveness.
In this paper, two networks are discussed: the Slovene Parliamentary Party Network~\cite{ferligoj1996analysis} and the Gahuku-Gama Subtribe Network~\cite{read1954cultures}.
The Slovene Parliamentary Party Network is a genuine alliance of political parties that depicts the connections between the parties in the Slovenian parliament in 1994. 
Slovene Parliamentary Party has 10 political parties: $\{$SKD, ZS, SDSS, ZLSD, DS, ZS-ESS, SNS, SLS, SPS-SNS$\}$. 
The cooperative relationship between political parties can be formalized into a network by quantifying the connections using seven levels of association, ranging from very hostile (-3) to very friendly (3).
Here, we change all negative elements of the adjacency matrix to $-1$ and all positive elements to $1$.
Thus, some of the parties will be grouped into the same clusters based on their mutual cooperation or antagonistic relationship.

The Gahuku-Gama Subtribes Network, established in 1954, encompasses 16 subtribes in the highlands of New Guinea. 
These subtribes are abbreviated as $\{$KOHIK, NAGAD, GAMA, GEHAM, ALIKA, NAGAM, GAHUK, GAVEV, KOTUN, OVE, MASIL, UKUDZ, NOTOH, ASARO, UHETO, SEUVE$\}$. 
Within this network, political alliances and class struggles exist among the 16 subtribes, highlighting interactions between them.
The relationships among the 16 subraces are represented by a weighted adjacency matrix. 
In this matrix, a value of $1$ signifies a political alliance, $-1$ represents a class struggle, and $0$ indicates that there is no specific relation between subraces.

\subsection*{Metric}
We present metrics to measure the balance of the signed network by counting the number of imbalanced circles~\cite{kirkley2019balance}.
The definition for this metric is as follows,
\begin{equation}\label{eqs:eq8}
	{B_s(z)=\frac{1}{4}log\frac{det[\alpha\lambda*I-(A^+-A^-)]}{det[\alpha\lambda*I-(A^++A^-)]}},
\end{equation}
where $\alpha$ is a constant and $\alpha\ge1$, $\lambda$ denotes the largest eigenvalues of $A^{+}+A^{-}$ and $A^{+}-A^{-}$.
In this paper, we employ $B_s$ to measure the structural balance of communities in the synthetic network and investigate the impact of structural balance on the motif-based method.
$B_s$ denotes the degree of imbalance in the structure of networks.
Thus, a smaller $B_s$ indicates a more balanced network, while a larger value suggests imbalance.

To assess the effectiveness of the motif-based method for community detection, we used Jaccard similarity to measure similarity between the detected community and a benchmark community structure~\cite{ferligoj1996analysis}. 
Jaccard similarity is a widely used metric that quantifies the overlap or agreement between two sets by calculating the ratio of the intersection to the union of the sets.
Two communities are represented as ${A_1, A_2, ..., A_K}$ and ${B_1, B_2, ..., B_K}$, where $K$ denotes the number of communities in networks A and B. 
Both networks share the same set of vertex $N$, which consists of the nodes present in both communities.
For each community, the similarity between the corresponding communities in networks A and B can be calculated as follows,
\begin{equation}\label{eqs:eq9}
	{s_{ij}=\frac{|A_i\bigcap B_j|}{|A_i\bigcup B_j|}}.
\end{equation}
To obtain the comprehensive performance, we average the similarity over each community, the corresponding formula is given in Eq.\ref{eqs:eq10},
\begin{equation}\label{eqs:eq10}
	{S=\frac{\sum\limits^{K}_{l=1}s_l}{K}}.
\end{equation} 

\section*{Results}
\subsection*{Synthetic networks}
In this section, we test the performance of the motif-based methods on synthetic networks with different network properties. 
By adjusting the number of links within communities and the ratio of negative links, we examine the performance of the motif-based method using the model $SN(4, 32, 32, p^{in}, 0, p^-_{in})$ from two perspectives: link density and structural balance. 
The parameter $p^{in}$ is varied from 0.1 to 0.9 with a step of 0.1.

\begin{figure}
	\centering
	\includegraphics[width =1.0 \textwidth]{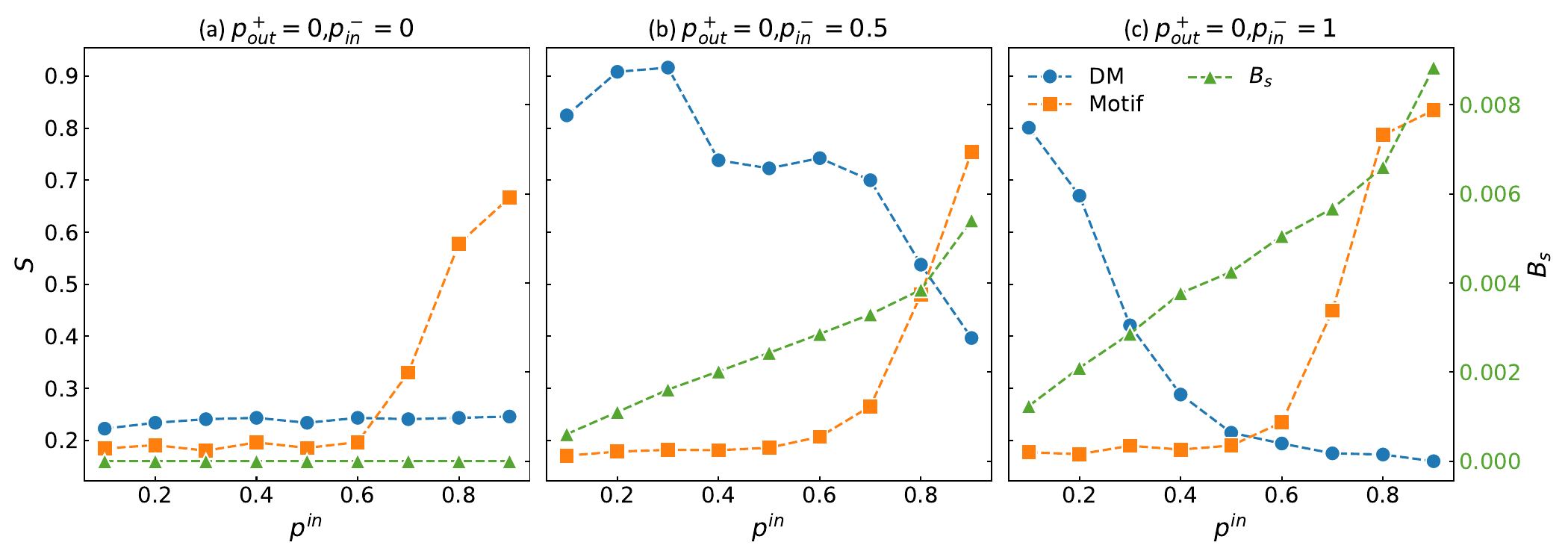}
	\caption{The Jaccard similarity and community balance plotted as a function of $p^{in}$ for various ratios of negative links within community on synthetic networks when $p^{+}_{out}=0$. 
		We compare the performance of DM with motif-based algorithms under different parameter settings.
		The synthetic network is generated by the SN(4, 32, 32, $p^{in}$, 0, $p_{in}^-$).
		(a) $p_{in}^-$= 0; (b) $p_{in}^-$=0.5; (c) $p_{in}^-$= 1}
    \label{fig:2}
\end{figure}

With higher values of $p^{in}$, there are fewer links between nodes in different communities, and the boundaries between communities become clearer, making it easier to distinguish them accurately. 
On the contrary, lower values of $p^{in}$ result in blurred boundaries between communities, and it is more difficult to identify the community.
In Figs.~\ref{fig:2}(a)-(c), we plot the similarity and network balance as a function of $p^{in}$ for various ratios of negative links within communities.
For a given $p_{in}^-$, it becomes apparent that the motif-based method exhibits a higher similarity to DM (a well-known community detection method in signed networks) as the value of $p^{in}$ increases.
The result suggests that the motif-based method has an advantage in distinguishable community structures.
For example, as can be seen from Fig.~\ref{fig:2}(c), the motif-based method performs better than the DM when $p^{in}$ is greater than 0.54.

In Figs.\ref{fig:2}(a)-(c), we present the performance of the motif-based method for different values of $p_{in}^-$, where $p_{in}^-= 0, 0.5$ and $1$.
As observed, the balance of the network, even for the same $p^{in}$, gradually worsens as $p^-_{in}$ increases. 
This is because the presence of more negative links within communities leads to a decrease in structural balance, making communities less balanced.
In such cases, the performance of the motif-based method remains robust and does not deteriorate when the density of links within communities is greater. 
In fact, the higher similarity observed for the motif-based method in Fig.\ref{fig:2}(c) compared to Fig.\ref{fig:2}(b) indicates that the method is remarkably robust under different levels of balanced communities. 
This advantageous characteristic allows the motif-based method to accurately detect communities even in networks with a lower structural balance.

\begin{figure}
	\centering
	\includegraphics[width =1.0 \textwidth]{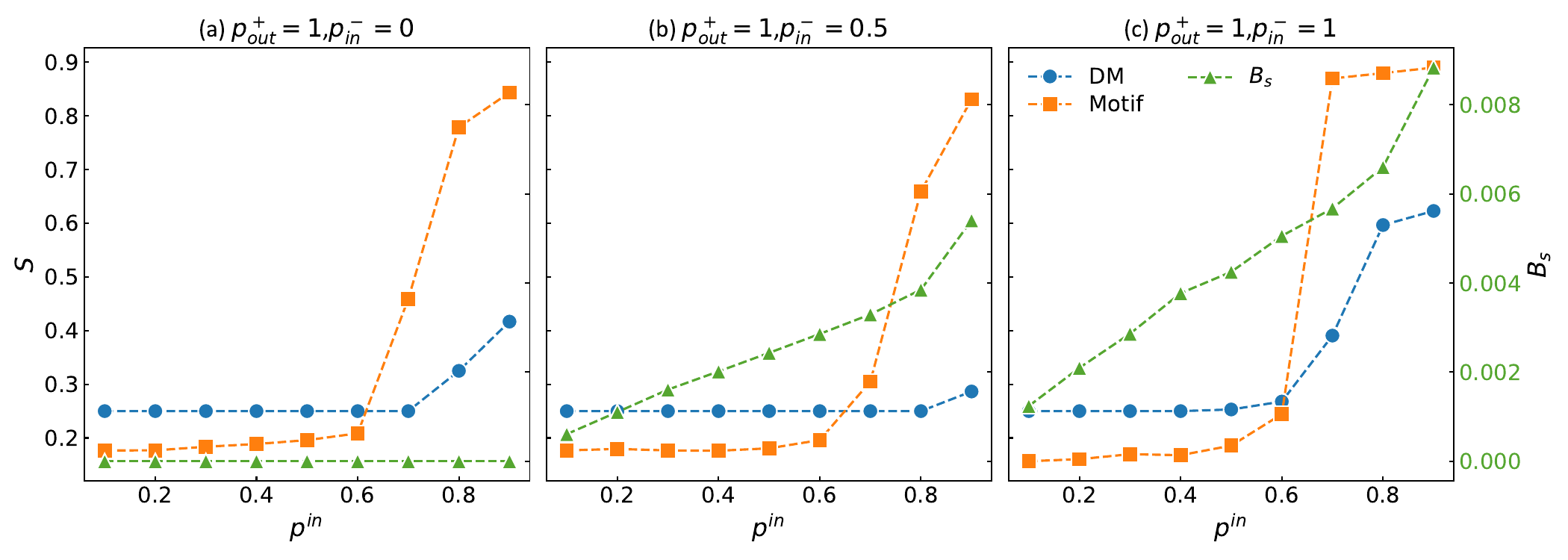}
	\caption{The Jaccard similarity and community balance plotted as a function of $p^{in}$ for different ratios of negative links on synthetic networks when $p^{+}_{out}=1$. 
		The synthetic network is generated by the SN(4,32,32,$p^{in}$,1,$p_{in}^-$). 
		(a) $p_{in}^-$= 0; (b) $p_{in}^-$=0.5; (c) $p_{in}^-$= 1}
    \label{fig:3}
\end{figure}

The above results are analysed on a signed network where all the links between communities are negative.
The task of community detection conducted on the signed network of the above configuration is relatively straightforward because clustering negative links tends to classify them as inter-community links.
This clear separation between communities based on negative links makes the detection process easier.
Next, we test the performance of the motif-based method on a network where all links between communities are positive. 
To create different networks for testing, we generate them using the $SN(4, 32, 32, p^{in}, 1, p^-_{in})$ model, which presents a greater challenge to the performance of community detection algorithms. 
We then compare the performance of the motif-based method with DM in a more challenging scenario, as shown in Fig.~\ref{fig:3}.
When the ratio of negative links within communities is relatively large, the motif-based method significantly outperforms DM, which is consistent with the findings observed in Fig.~\ref{fig:2}. 
The results indicate that the motif-based method still performs well even when there are some positive links between communities. 

By combining the findings from Fig.~\ref{fig:2} and \ref{fig:3}, a consistent conclusion can be drawn: the performance of the motif-based method in weakly balanced communities is not inferior to that in balanced communities, especially when the community structure is clearly distinguishable. 
This suggests that the motif-based method exhibits robustness and effectiveness in community detection across different types of community structures, including both balanced and weakly balanced communities.
The main reason behind this finding is that the density of links within communities plays a crucial role in determining the performance of the motif-based method. 
On the other hand, we observe the effect of $p_{out}^+$ while keeping $p_{in}^-$ constant~(e.g. Figs.~\ref{fig:2}(a) and \ref{fig:3}(a)) on the performance of the motif-based method.
In fact, it shows the lower similarity in Fig.~\ref{fig:2} compared to Fig.~\ref{fig:3} when $p^{in}$ is larger, suggesting that our method performs better in networks with a very clear community and an imbalanced structure.

\begin{figure}
	\centering
	\includegraphics[width =1.0 \textwidth]{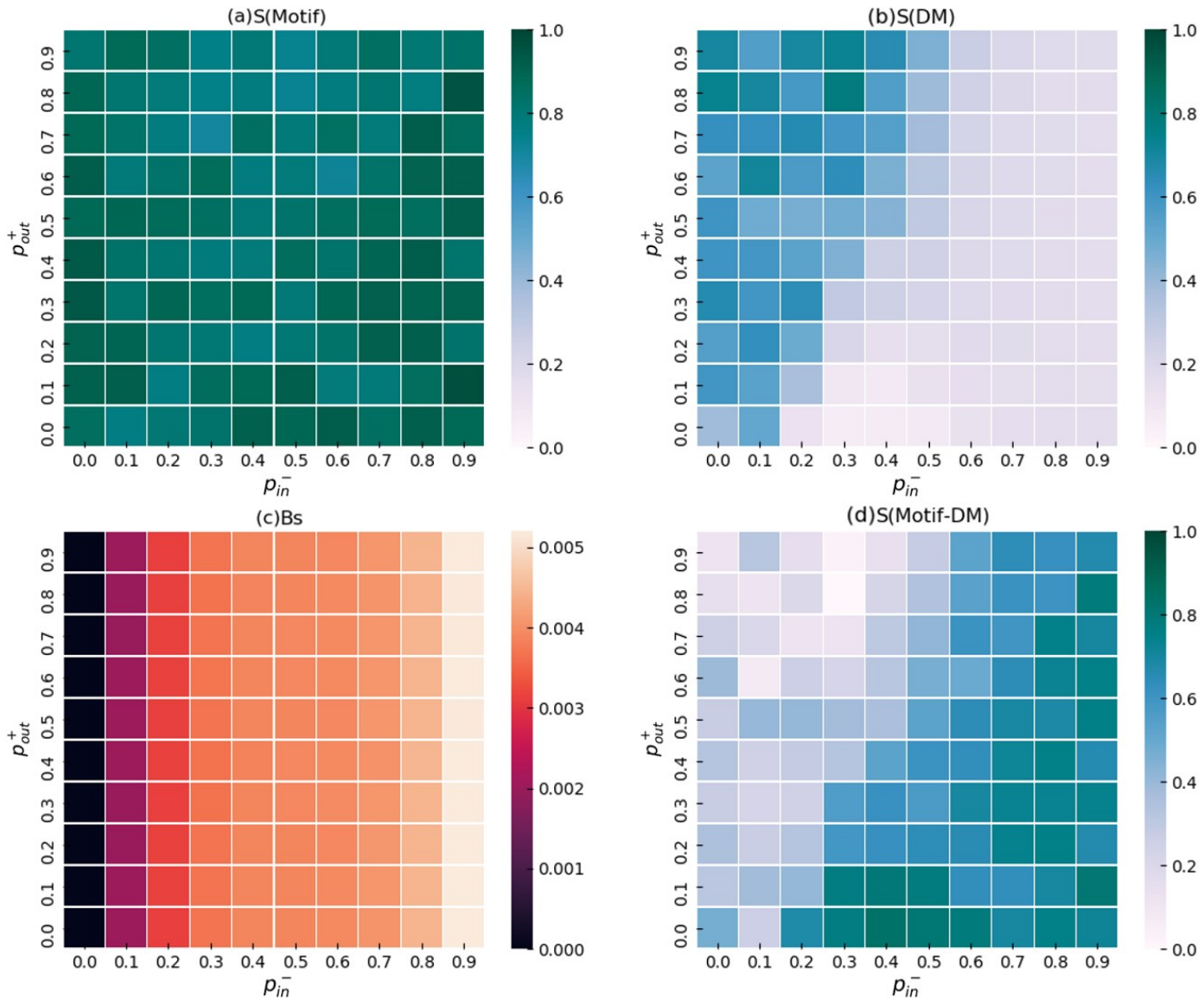}
	\caption {The similarity of both motif-based method and DM as well as the balance of benchmark communities under different signed parameters ($p^-_{in}$ and $p^+_{out}$) with a given $p^{in}=0.9$. 
		(a)(b) The similarity between the detected community and benchmark community for both method under various $p^-_{in}$ and $p^+_{out}$.
		(c) Balance $B_s$ of benchmark community.
		(d) Difference of similarity between the motif-based method and DM.
		Darker colours denote that our method has a more pronounced advantage over DM.
        }
        \label{fig:4}
\end{figure}

To further analyse the impact of the signed information configuration, specifically various values of $p^-_{in}$ and $p^+_{out}$, on the performance of the motif-based method, we keep $p^{in}$ fixed. 
We then present the similarity of different methods and the balance of benchmark communities under different parameter settings. 
This analysis allows us to understand how changes in the ratio of negative links within communities and positive links between communities influence the community detection results and the overall balance of the identified communities.

Figure 4(a) shows that the similarity of the motif-based method exhibits subtle differences for different pairs of $p^-_{in}$ and $p^+_{out}$, which in turn determine the balance of the community structure. 
The performance of the motif-based method remains relatively consistent across various combinations of these parameters, indicating its robustness in handling different configurations of signed information.
On the contrary, in Fig.~\ref{fig:4}(b), we observe that the performance of the DM is more sensitive to the configuration of signed information, particularly in relation to the parameter $p^-_{in}$, which determines the balance of the community structure. 
In Fig.~\ref{fig:4}(c), we perform further analysis of community balance for different combinations of $p^-_{in}$ and $p^+_{out}$. 
The results show that as $p^-_{in}$ increases, the balance of communities gradually deteriorates.
The density of positive links between communities, determined by $p^+_{out}$, is not part of the communities themselves and therefore does not contribute to the balance of the communities.
In Fig.~\ref{fig:4}(d), as previously suggested, our motif-based method significantly outperforms the DM method when the community structure is very clear.
Furthermore, as $p_{in}^-$ increases, the advantage of the motif-based method becomes even more evident. 
This observation indicates that our motif-based approach is particularly effective in scenarios where there are well-defined community structures, highlighting its robustness and advantage over the DM method in handling such configurations of signed networks.

\subsection*{Real-world networks}

\begin{figure}
	\centering
	\includegraphics[width =1.0 \textwidth]{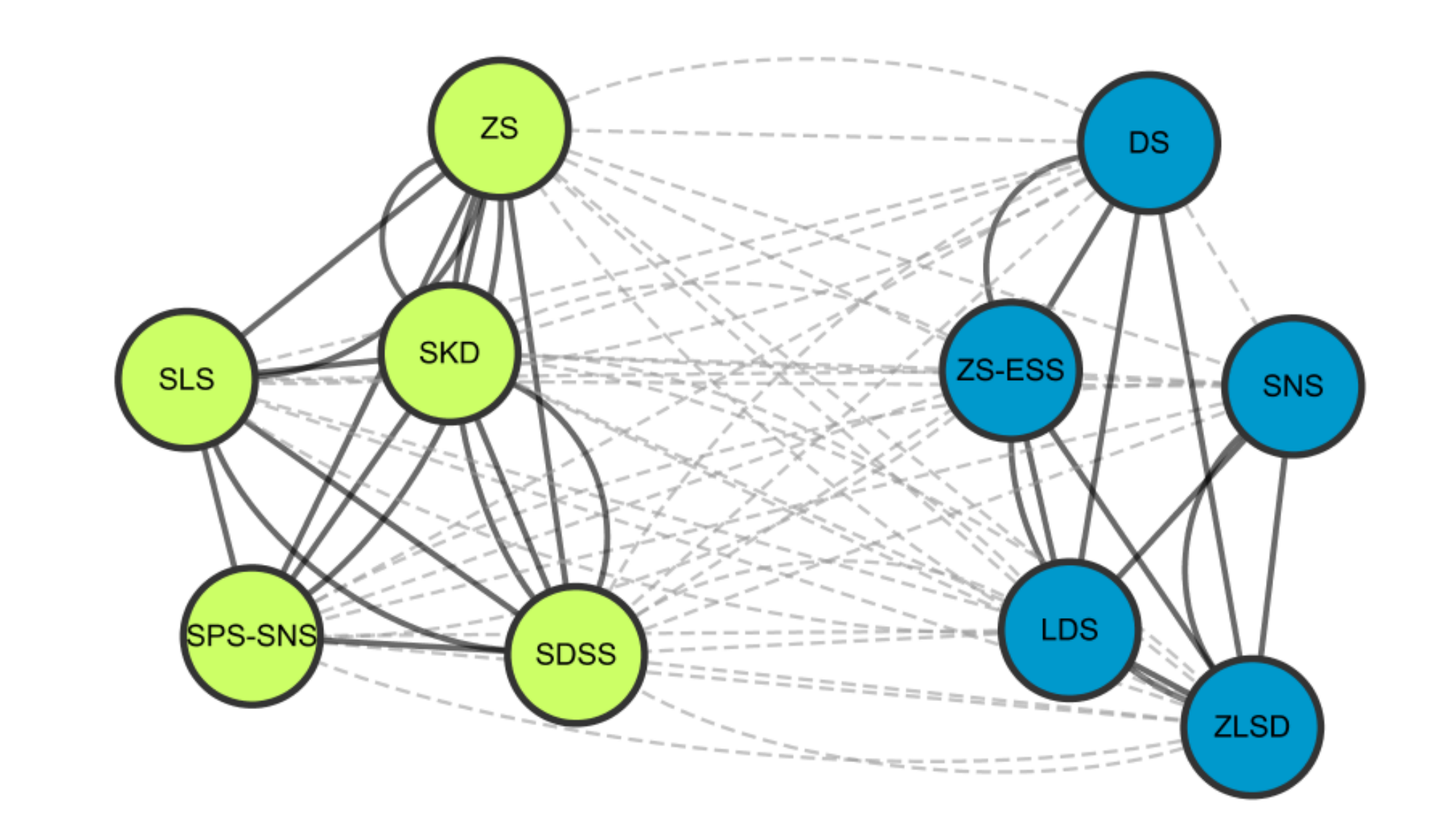}
	\caption{The detection community of Slovene Parliamentary Party Network, where the solid and dotted line represents the positive and the negative link, respectively.}
         \label{fig:5}
\end{figure}

\begin{figure}
	\centering
	\includegraphics[width =1.0 \textwidth]{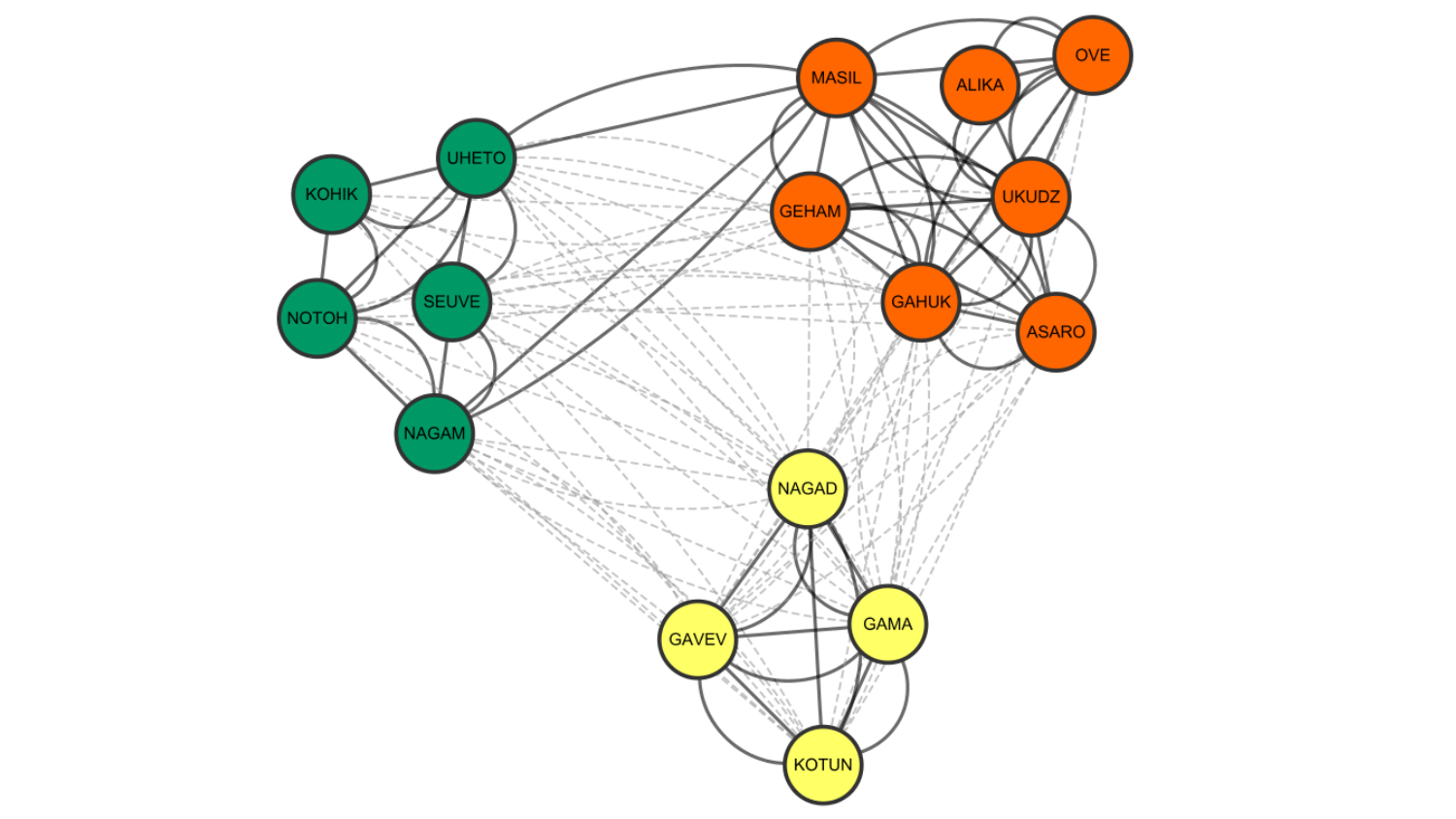}
	\caption{The detection community of Gahuku-Gama Subtribes Network, where the solid and dotted line represents the positive and the negative link, respectively.}
         \label{fig:6}
\end{figure}

We applied the motif-based method to real networks, specifically the Slovene Parliamentary Party and the Gahuku-Gama Subtribes Network, and obtained highly accurate results. 
As shown in Figs.~\ref{fig:5} and \ref{fig:6}, the detected communities for the Slovene Parliamentary Party, that is, $\{$ SDSS, SKD, ZS, SLS, SPS-SNS$\}$ and $\{$ DS, ZS-ESS, SNS, LDS, ZLSD $\}$, precisely corresponded to the real communities, confirming the effectiveness of our approach in identifying real-world communities.
In particular, the motif-based method e xhibited robustness in handling the configuration of signed links, where it fully utilised the signed information to detect communities.
For instance, in the case of the Slovene Parliamentary Party, when we removed the signed information from the links, the network degenerated into a fully connected one, making it impossible to detect communities using purely structural information.
However, our motif-based method, incorporating signed information, successfully identified the communities, which highlights the significant role of signed information. 
Similarly, for the Gahuku-Gama Subtribes Network shown in Fig.~\ref{fig:6}, the motif-based method accurately identified three communities: $\{$GEHAM, UKUDZ, NAGAD, NAGAM, ASARO, SEUVE, GAVEV$\}$, $\{$UHETO, GAHUK, KOHIK, OVE$\}$, and $\{$GAMA, ALIKA, KOTUN, MASIL, NOTOH$\}$, which closely aligned with the ground truth communities. 
These findings confirm the applicability of the motif-based method for community detection in real networks.

\section*{Discussion}
The signed network can describe positive and negative relations between users. 
Such a rich attitude information is crucial to uncovering the complex community structure in online social networks. 
Existing related works mainly focus on minimizing the proportion of negative edges within communities and dividing the links as inter-community connections.
However, negative links are abundant within communities in social networks and play an important role in the stability of community formation. 

In this work, we generalize the traditional concept of communities in signed networks considering the structural balance within communities and the density of connectivity within them.
On the basis of this, we propose a motif-based method to quantify the contribution of single links to the local network structure and identify communities.
We evaluated the performance of our algorithm and compared it with the DM algorithm on the signed synthetic network model. 
The result revealed that when the community structure is distinguishable, our algorithm significantly outperformed the DM algorithm even though the network is imbalanced. 
Furthermore, our algorithm exhibited a high level of robustness, being insensitive to sign configuration, indicating that its performance is independent of network balance.
On two real-world networks, our algorithm accurately identified the actual communities, demonstrating its excellent practical applicability. 
This work provides a new perspective for community detection in signed networks, calling for more efforts to identify community structures from the perspectives of community formation and balance, which lays the foundation for further related analytical work.

\section*{Acknowledgements}
Leyang Xue acknowledge the support of the China Scholarship Council Program.

\bibliography{reference}
\end{document}